# An Analytic Linear Accelerator Source Model for Monte Carlo Dose Calculations. I. Model Representation and Construction

**Zhen Tian[1], Yongbao Li[1,2], Michael Folkerts[1], Feng Shi[1], Steve B. Jiang[1], Xun Jia[1]**

[1]Department of Radiation Oncology, University of Texas Southwestern Medical Center, Dallas, TX 75390, USA
[2]Department of Engineering Physics, Tsinghua University, Beijing, 100084, China

Emails:     zhen.tian@utsouthwestern.edu,     steve.jiang@utsouthwestern.edu,
xun.jia@utsouthwestern.edu

Monte Carlo (MC) simulation is commonly considered as the most accurate method for radiation dose calculations. Accuracy of a source model for a linear accelerator (linac) is critical for the overall dose calculation accuracy. In this paper, we presented an analytical source model that we recently developed for GPU-based MC dose calculations. A key concept called phase-space-ring (PSR) was proposed. It contained a group of particles that were of the same type and close in energy and radial distance to the center of the phase-space plane. The model parameterized probability densities of particle location, direction for and energy for each primary photon PSR, scattered photon PSR and electron PSR. For a primary photon PSRs, the particle direction is assumed to be from the beam spot. A finite spot size is modeled with a 2D Gaussian distribution. For a scattered photon PSR, multiple Gaussian components were used to model the particle direction. The direction distribution of an electron PSRs was also modeled as a 2D Gaussian distribution with a large standard deviation. We also developed a method to analyze a phase-space file and derive corresponding model parameters. To test the accuracy of our linac source model in terms of representing the reference phase-space file, dose distributions of different open fields in a water phantom were calculated using our source model and compared to those directly calculated using the reference phase-space file. The average distance-to-agreement (DTA) was within 1 mm for the depth dose in the build-up region and beam penumbra regions. The root-mean-square (RMS) dose difference was within 1.1% for dose profiles at inner and outer beam regions. The maximal relative difference of output factors was within 0.5%. Good agreements were also found in an IMRT prostate patient case and an IMRT head-and-neck case. These results demonstrated the efficacy of our source model in terms of accurately representing a reference phase-space file.

**1. Introduction**

Monte Carlo (MC) method for dose calculation is desired in radiation therapy due to its well accepted accuracy (Rogers and Mohan, 2000; Keall *et al.*, 2000). Computational efficiency has been a major concern that prevents its clinical applications. Recently, there have been a lot of research interests in developing fast MC dose calculation methods on a graphics processing unit (GPU) platform. By utilizing the rapid parallel processing capability of a GPU and designing GPU-friendly parallelization schemes, tremendous acceleration factors have been observed compared to conventional CPU-based calculations. Dose calculations for a typical photon treatment plan can be completed in only tens of seconds (Jia *et al.*, 2010; Jia *et al.*, 2011; Hissoiny *et al.*, 2011; Jahnke *et al.*, 2012; Townson *et al.*, 2013).

Clinical applications of these MC dose engines require accurate source modeling for linear accelerators (linacs). Although previous studies (Jia *et al.*, 2010; Jia *et al.*, 2011; Hissoiny *et al.*, 2011; Jahnke *et al.*, 2012) have demonstrated accuracy of particle transport inside a patient body, source modeling in those GPU-base MC codes has been rarely reported. Recently, a phase-space based source model was developed for the package gDPM (Jia *et al.*, 2010; Jia *et al.*, 2011; Townson *et al.*, 2013). A key concept in this method was phase-space-let (PSL), which was generated by splitting a phase-space file into small pieces according to particle type, location, and energy. The PSL technique had several favorable features. When calculating dose in a treatment plan, it allowed only loading particles that were inside or close to jaw opening area, which substantially improved efficiency comparing to loading all particles from a huge phase-space file. In addition, it enabled an automatic commissioning process (Tian *et al.*, 2014), where weighing factors associated to PSLs were adjusted to finely tune the contributions of particles from them to accurately represent a linac beam. Despite these advantages, this PSL-based source model was less optimal for GPU platform. It usually took tens of seconds to prepare for the simulations, including loading a large amount of particle data from a hard drive to CPU memory, manipulating them for dose calculations, and transferring them to GPU memory. While time of tens of seconds was relative short for CPU-based MC simulations, it constituted a large portion of the total computational time for GPU-based MC dose calculations, posing a bottleneck to further improve efficiency.

In this regard, an analytical source model is more preferred for GPU-based dose engines, in that data preparation part can be avoided. Over the years, source models of this type have been extensively studied (Ma *et al.*, 1997; Ma, 1998; Deng *et al.*, 2000; Jiang *et al.*, 2000; Davidson *et al.*, 2008; Fix *et al.*, 2004; Verhaegen and Seuntjens, 2003). Generally speaking, an analytical source model contains multiple beam components. Particle energy, location, and direction distributions are explicitly expressed for each component via analytical functions. When conducting a dose calculation, source particles from a linac can be sampled from these probability distributions on the fly. As the sampling tasks are usually light weighted, analytical source models are in principle favorable for a GPU-base MC engine and are expected to outperform phase-space file based models in terms of efficiency.

A good analytical source model should have the following features: (1) The model should be able to represent the actual particle distributions from a linac beam; (2) It should be easy to derive parameters in the model; (3) The model should be simple enough to allow an easy sampling procedure of source particles during dose calculations; (4) The model should be commissionable to a real clinical beam with an automated commissioning process. Furthermore, in the context of GPU-based dose calculations, it is preferred to develop a sampling approach for the source model to coordinate the sampling processes among different GPU threads in order to be maximally compliant with the GPU's single-instruction multiple data (SIMD) scheme (Jia *et al.*, 2014) and to utilize the available processing capacity. Otherwise, the overall efficiency can be easily impaired, as sampling processes on different threads may run into different sub-sources, which yield execution divergence among them causing the so-called thread-divergence problem.

Aiming at these objectives, we have developed an analytical source model of photon beams for a GPU-based MC dose engine. We will present various aspects of this model in a series of two papers. In this current paper, we will present the model itself, in particular, how different components in the model are parameterized analytically. We will also derive parameters in our model by analyzing a phase-space file. Due to page limitations, we will present in a separate paper details about a GPU-friendly sampling strategy, as well as a beam commissioning method (Tian *et al.*, 2015a).

## 2. Methods and Materials

### 2.1 Phase-space ring (PSR) concept

In our source model, we would like to parameterize particle distributions on a phase-space plane that is $z_{phsp}$ away from the target. This phase-space plane resides above the upper jaws in order to accommodate different jaw settings in real treatment plans. The beam on and above this plane is assumed to be rotationally symmetric due to the symmetric hardware geometry of the linac machine head above jaws. We introduce a concept called phase-space ring (PSR). It refers to a group of particles that are of the same type, have the same interaction history (primary or secondary), reside in a narrow ring in the phase-space plane, and are in a certain energy range. Fig. 1(a) illustrates the partition of the phase-space plane into rings, whereas the partition according to energy and interaction history is not explicitly shown. Each PSR is associated with three indices denoted as $PSR_{stype,Rbin,Ebin}$. $stype$ is sub-source index with $pp$, $ps$, and $e$ corresponding to the three sub-sources we consider, namely primary photon, secondary photon and electron sub-sources. $Rbin = 0, 1, …, NR$ and $Ebin = 0, 1, …, NE$ are indices for the ring and energy bins, respectively. Our model will parameterize probability distributions for each of the PSRs.

The reason for introducing this PSR concept is to employ the beam rotational symmetry, particularly for two purposes. First, we will later construct our model by analyzing particles in a phase-space file that we want to replicate with our model, referred as "a reference phase-space file". The PSR concept with an inherent assumption of rotational symmetry allows us to use all particles in a ring to build a model to reduce statistical uncertainty. Second, we will also develop an automatic commissioning method to account for the difference between the model built based on the phase-space file and a real linac beam (Tian *et al.*, 2015a). PSR naturally integrates the rotational symmetry in the commissioning process. Comparing with our previous PSL-based approach, using PSR avoids adding regularization terms to the commissioning model (Townson *et al.*, 2013; Tian *et al.*, 2014).

In our model, primary and scattered photons were treated separately for the following reasons. 1) Primary and scattered photons have distinct directional distributions, requiring modeling them separately. 2) Separating primary and scattered photons allow us to determine different sampling areas for the two types of PSR for a given jaw opening in a treatment plan. It will improve the computational efficiency of dose calculations, which will be discussed in (Tian *et al.*, 2015a). 3) Modeling these two types of photon PSRs separately permit sampling primary photon and scattered photon separately in dose calcualtions (Tian *et al.*, 2015a), which could alleviate the GPU thread divergence issues. 4) In the automatic commissioning process, separate weighting factors can be applied to the primary and the scattered photon components, leading to more degrees of freedom to commission a beam.

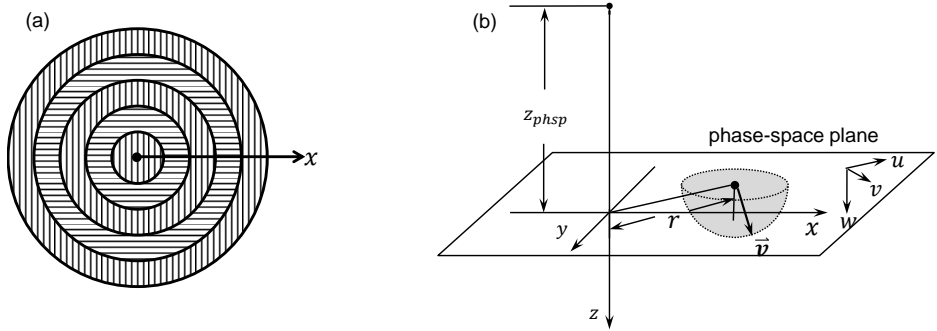

**Figure 1.** (a) Illustration of phase-space ring (PSR) concept. (b) The coordinate system defined for particle direction in our PSR-based source model.

### 2.2 General model structure

Based on the idea of the PSR concept, we start parameterizing the distribution of particles in each PSR sub-source. Let us start with the definition of our coordinate system. As illustrated in Fig. 1(b), at each point on the phase-space plane, the particle direction vector $\vec{v}$ is on a hemisphere surface, ignoring back-scattered photons that travel upwards. We define the component $u$ for the particle direction to be along the radial direction. The component $v$ is in the plane but perpendicular to $u$. The third component $w$ is perpendicular to the phase-space plane, satisfying $= \sqrt{1 - u^2 - v^2}$. Let us consider the probability density function $P_{stype,Rbin,Ebin}(r, u, v, E)$, namely the probably density for a sub-source indexed by $stype, Rbin, Ebin$ to have a particle with a radius $r$, a direction $(u, v)$ and energy $E$. Note that the third component $w$ is not explicitly written due to the constraint of unit vector length. The probability density function can be generally expressed as:

$$P_{stype,Rbin,Ebin}(r, u, v, E)$$

$$= W_{stype,Rbin,Ebin}\, p_{stype,Rbin,Ebin}(r, u, v, E). \tag{1}$$

The first term on the right side $W_{stype,Rbin,Ebin}$ denote the relative probability among different PSRs. It is normalized in the sense that

$$\sum_{stype} \sum_{Rbin} \sum_{Ebin} W_{stype,Rbin,Ebin} = 1. \tag{2}$$

For the second term on the right hand side $p_{stype,Rbin,Ebin}(r, u, v, E)$, it is further broken down to

$$p_{stype,Rbin,Ebin}(r, u, v, E) = \pi_{Rbin}(r)\pi_{Ebin}(E)p_{stype,Rbin,Ebin}(u, v|r, E), \tag{3}$$

where $\pi(.)$ is a uniform distribution for the radius or the energy in the corresponding bins. The third term is the condition probability density of particle direction given its energy and radius, whose expression varies depending on the sub-source type, which will be presented in section 2.3.

### 2.3 Model construction

### 2.3.1 Splitting reference phase-space file into PSRs

To derive $W_{stype,Rbin,Ebin}$ and $p_{stype,Rbin,Ebin}(u, v|r, E)$ in our model from a reference phase-space file, we process the phase-space file through the following steps.

After projecting particles in the reference phase-space file to our phase-space plane at $z_{phsp}$, we would like to separate the files into three sub-sources. While electron and photons are easily identified based on particle type, separating primary and scattered photons is based on geometry considerations. Specifically, we backproject each particle to the level of the electron target. If the photon is within a circular area defined by the radius of the electron beam, it is considered as the primary photon. This radius of the electron beam is usually recorded in the header of the phase-space file. We remark that for the EGSnrc phase-space files, a LATCH variable in each particle may be used to to label its interaction history, which makes this step straightforward. However, in the IAEA phase-space files (Capote, 2007) we used, such a variable is not available. We hence separated photons in the two groups by analyzing the geometry.

The next step is to perform a coordinate transformation. The reference phase-space file uses a Cartesian coordinate system such that the three axis of $(x, y, z)$ are parallel to the direction $(u_p, v_p, w_p)$, respectively. Assuming the rotational geometry, the azimuthal angle becomes irrelevant, and hence we compute the coordinate $r = \sqrt{x^2 + y^2}$ for each particle. Meanwhile, we also compute the particle direction vector in the coordinate $(u, v, w)$ in our system, where $u$ is along the radial. Specifically, this can be achieved by $u = u_p \cos\theta + v_p \sin\theta$ and $v = -u_p \sin\theta + v_p \cos\theta$, where $\theta = \arctan\left(\frac{y}{x}\right)$ is the azimuthal angle.

After these preprocessing procedures, all the source particles in the file can be distributed into different PSRs based on their types (photon or electron), interaction histories (primary or secondary), energy values, and radial distances.

### 2.3.2 Model construction

#### 2.3.2.1 Weighting factors

After splitting the phase-space file into different PSRs, we are ready to analyze their statistical properties to build models for them. First, the weighting factor of each PSR $W_{stype,Rbin,Ebin}$ in Eq.(1) is proportional to the number of particles in it:

$$W_{stype,i} = \frac{NP_{stype,i}}{\sum_{stype}\sum_i NP_{stype,i}},\qquad(4)$$

where the subscript $i = 1, \ldots, NR \times NE$ is a short notation for $Rbin, Ebin$ from hereon. $NP_{stype,i}$ is the number of particles for a specific PSR.

#### 2.3.2.2 Primary photons

We then need to construct the probability density of particle direction $p_{stype,i}(u,v|r,E)$ in Eq.(3) conditioned on the particle's energy and radius, and to derive the parameters based on the reference phase-space file. For a primary photon, its direction is determined once its radius $r$ is determined, because it comes from the source directly. To account for the finite spot size, we assume that the source position follows a 2-D Gaussian distribution with a zero mean value and a standard deviation $\sigma_s$. Thus the direction distribution of primary photons is formulated as

$$p_{pp,i}(u,v|r,E) = \int \delta(u-u_0)\delta(v-v_0)\frac{1}{2\pi\sigma_s^2}e^{-\frac{x_s^2+y_s^2}{2\sigma_s^2}}\,dx_s dy_s,\qquad(5)$$

where $(x_s, y_s)$ is a location for the initial photon position on the initial source plane and $\delta(.)$ is the delta function. With this initial position of a primary photon, its particle direction can be expressed as

$$u_0 = \frac{r-x_s}{\sqrt{(r-x_s)^2+y_s^2+z_{phsp}^2}},$$
$$v_0 = \frac{-y_s}{\sqrt{(r-x_s)^2+y_s^2+z_{phsp}^2}}.\qquad(6)$$

See Fig. 2(a) for the illustration of the geometry.

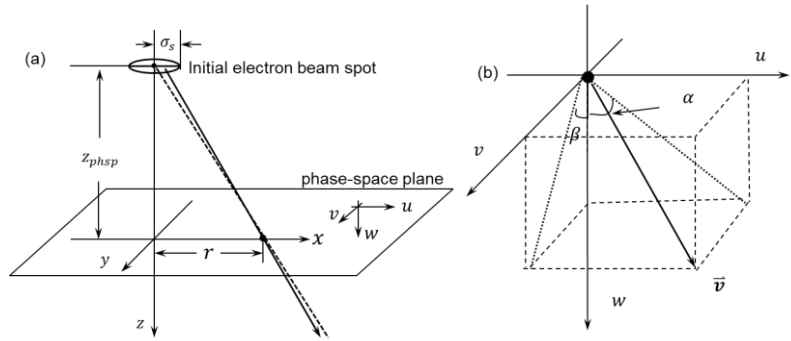

**Figure 2.** (a) Illustration of the geometry for primary photons. (b) Definition of angles used in scattered photons.

The only parameter in this primary photon PSR is the standard deviation $\sigma_s$. It is a tuning parameter of our model to control the impacts of the spot size, which affects the gradient of beam penumbra and output factors, particularly for small open fields. We remark that this parameter is not necessarily the same as the beam spot size specified in the header file for the reference phase-space file.

### 2.3.2.3 Scattered photons

For scattered photons, the particle direction can be in general arbitrary on the hemisphere shown in Fig. 1(b). Note that we ignored the back-scattered photons that travel upwards. To get insights about the direction distributions of scattered photons, we analyzed a phase-space file for a Varian TrueBeam linac (Varian Medical System, Palo Alto, CA) and plotted that for a few PSRs on a hemisphere. We had the following observations depicted in Fig. 3. 1) The directions of scattered photon clustered in a certain region. The width of the cluster depends on both PSR energy and ring radial indices; 2) The center of the region along the $u$ direction is related to the PSR's radius. The larger the radius is, the further the center is away from $u = 0$; 3) The distributions are symmetric about the $v$ axis, which comes from the rotational symmetry of the beam.

These observations motivated us to fit each distribution by a function form of multiple 2D Gaussian functions. However, if we were to simply use Gaussian functions on variables $u$ and $v$, the sampled results in an MC simulation would be sometimes unphysical with $u^2 + v^2 > 1$. To avoid this issue, we switch to represent the particle direction vector using two angle variables, $\alpha = \text{atan}\left(\frac{u}{w}\right)$ and $\beta = \text{atan}\left(\frac{v}{w}\right)$, as illustrated in Fig.2 (b). The direction distribution model for each scattered photon PSR can be formulated as

$$p_{ps,i}(\alpha, \beta | r, E) = \sum_{k=1}^{K} G_{i,k}(\alpha, \beta) \tag{7}$$

$$= \sum_{k=1}^{K} \frac{c_{i,k}}{2\pi \times \sigma_{\alpha,i,k} \times \sigma_{\beta,i,k}} e^{-\frac{(\alpha - \mu_{i,k})^2}{2\sigma_{\alpha,i,k}^2} - \frac{\beta^2}{2\sigma_{\beta,i,k}^2}},$$

where $c_{i,k}$ denotes the relative weight of the $k_{th}$ 2D Gaussian component among the multiple Gaussian distributions for the $i_{th}$ scattered PSR. They should satisfy $c_{i,k} \geq 0$ and $\sum_k c_{i,k} = 1$. $\sigma_{\alpha,i,k}$ and $\sigma_{\beta,i,k}$ are standard deviations in the $\alpha$ and $\beta$ dimensions, respectively. $\mu_{i,k}$ denotes the mean value in the $\alpha$ dimension to model the shift of the distribution center along the $u$ direction. The mean value in $\beta$ dimension is assumed to be 0 due to our symmetric assumption. Once the two angles $\alpha$ and $\beta$ are sampled, $(u, v)$ can be computed as

$$u = \frac{\tan \alpha}{\sqrt{1 + (\tan \alpha)^2 + (\tan \beta)^2}}, \tag{8}$$
$$v = \frac{\tan \beta}{\sqrt{1 + (\tan \alpha)^2 + (\tan \beta)^2}}.$$

Each Gaussian component of one PSR model has four unknown parameters, $c_{i,k}$, $\sigma_{\alpha,i,k}$, $\sigma_{\beta,i,k}$ and $\mu_{\alpha,i,k}$. To derive these parameters from the reference phase-space file, we further construct a histogram of scattered photons for a given PSR as a function $\alpha$ and $\beta$. A nonlinear curve fitting toolbox with Levenberg-Marquardt algorithm (Moré, 1978) in MatLab (MathWorks, Natick, MA) was used to fit the histogram in the analytical form of Eq. (7), yielding the parameters for this PSR. The number of Gaussian components $K$ is a tuning parameter in our model. An empirically chosen value of $K = 3$ were found sufficient.

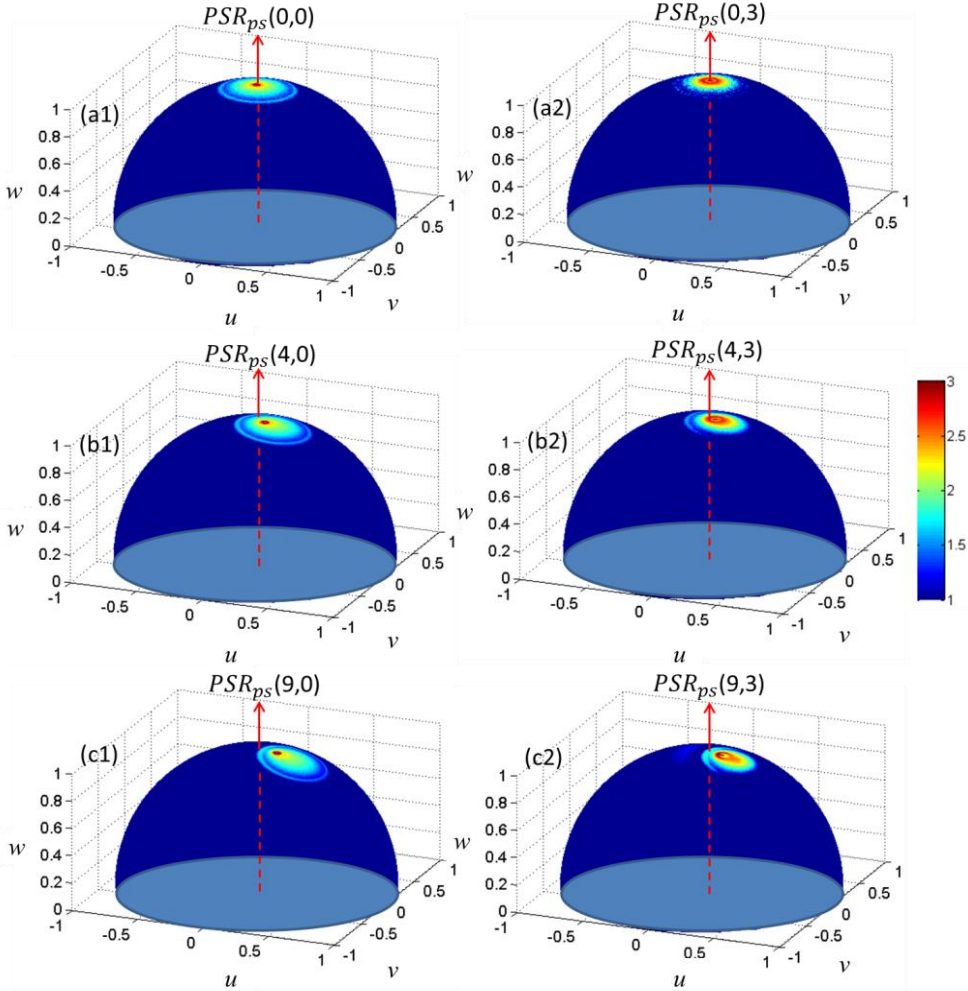

**Figure 3.** Illustration of direction distributions of the scattered photons in $PSR_{ps}(Rbin, Ebin)$. Three rows with *Rbin=0,4,9* are for rings with radius ranging from 0~0.4 cm, 1.6~2.0 cm, and 3.6~4.0 cm respectively; two columns with *Ebin=0,3* are for two energy bins with energy ranging from 0~0.636 MeV and 1.908~2.544 MeV. The red arrow indicates the direction *(u,v,w)=(0,0,1)*. Color of the sphere surface represents the logarithm of the probability density.

One issue worth mentioning is the underestimation of scattered photon probability at the primary photon direction during the phase-space file splitting. In fact, because of splitting primary and scattered photons based on geometry analysis, we cannot separate them if they are along the same direction. In our implementation, we assumed all the photons coming from the direction of the source spot are primary photons. This yielded a hollow region in the histogram for the direction of the scattered photons. To address this issue and still fit the scattered distribution with the Gaussian model, we only used the histogram data outside the hollow region for model fitting. After that, we computed particle numbers predicted by the Gaussian model in the hollow region. The difference between this number and that in the histogram is the particle number that was underestimated when splitting the phase-space file into primary and scattered photons.

We hence removed this number of photons from the corresponding primary PSR and add it back to the scattered photon PSR. Note that the PSR weighting factors were computed in Eq. (4) using the corrected particle numbers.

### 2.3.2.4 Electron PSR

Contaminant electrons constitute the third component $PSR_e$ that we considered in our analytic source model. The amount of the contaminant electrons contained in a reference phase-space file is usually too small to allow a good statistical analysis on the direction distribution. However, when plotting the particle directions in Fig. 4, it seems that the distribution is still Gaussian-like.

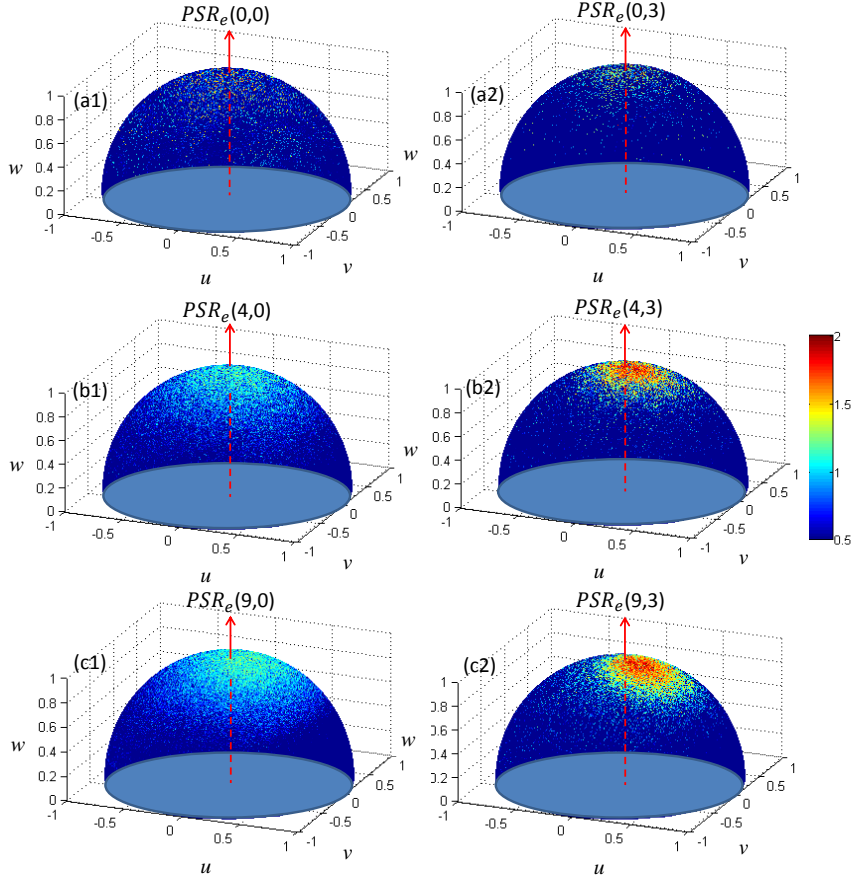

**Figure 4.** Illustration of direction distributions of the contaminant electrons in $PSR_e(Rbin, Ebin)$. Three rows with *Rbin=0,4,9* are for rings with radius ranging from 0~0.4 cm, 1.6~2.0 cm, and 3.6~4.0 cm respectively; two columns with *Ebin=0,3* are for two energy bins with energy ranging from 0~0.636 MeV and 1.908~2.544 MeV. The red arrow indicates the direction *(u,v;w)=(0,0,1)*. Color of the sphere surface represents the logarithm of the probability density.

Hence we assume that the direction distributions of electrons for all $PSR_e$ is a 2D Gaussian distribution with two large standard deviations $\sigma_{e\alpha}$ and $\sigma_{e\beta}$:

$$p_{e,i}(\alpha, \beta | r, E) = \frac{1}{2\pi \times \sigma_{e\alpha} \times \sigma_{e\beta}} e^{-\frac{\alpha^2}{2\sigma_{e\alpha}^2} - \frac{\beta^2}{2\sigma_{e\beta}^2}}. \tag{9}$$

Again, the direction $(u, v)$ is related to the two angles $\alpha$ and $\beta$ via Eq. (8). Although it was observed that the center of the electrons' direction distribution also shifted along the $u$ axis for the PSRs with large radius, this was ignored in the source modeling to keep our source model simple. In addition, those two standard deviations were assumed to be independent of the PSR index $i$ and empirically set to be $\pi/6$. This approximate model may not exactly replicate the direction distribution of the contaminant electrons contained in the reference phase-space file. However, these electrons account for ~1% of the particles in the file and only contribute to dose in the shallow depths that is usually not of interest for photon beams. Besides, our experimental results shown later illustrate that this approximate electron model could provide clinically acceptable accuracy.

*2.4 Materials*

The main purpose of this paper is to introduce our new PSR-based analytical source model and the method to analyze a reference phase-space file to derive model parameters. Due to space limit, the details about efficient integration of our source model into a GPU-based dose engine, as well as an automatic commissioning approach, will be presented in a separate manuscript (Tian *et al.*, 2015a). As such, we only focus on demonstrating the efficacy of our source model in this paper, specifically, in terms of its capability to accurately represent a reference phase-space file. This will be achieved by comparing doses calculated using our analytical source model and the dose calculated using the reference phase-space file.

We have built our PSR-based analytical source model for a Varian (Varian Medical System, Palo Alto, CA) TrueBeam 6MV beam. A set of 50 phase-space files for this beam provided by the manufacture was used as the reference phase-space file to construct our model and derive its parameters. In these files, there were about 2.5 billion particles (~ 50 GB data size). There particles were projected to a phase-space plane $z_{phsp} = 26.7\ cm$ from the target, which is above the upper jaws. Flattening filter was the last component simulated when generating these phase-space files using the MC method for linac head simulation. In our PSR model, there were ten energy bins with a resolution 0.636 MeV (one tenth of the maximal particle energy in the phase-space files). Along the radial direction, there are 20 rings with a resolution of 0.4 cm.

The dose engine we used in this paper is our newly developed MC dose calculation package under OpenCL environment (Tian *et al.*, 2015b). The use of OpenCL allows the code to run on CPU, GPUs from different vendors, and even heterogeneous platforms (Khronos OpenCL Working Group, 2013). The accuracy of this new dose engine has been demonstrated to be within 0.53% of our previously developed GPU-based dose engine gDPM (Jia *et al.*, 2010; Jia *et al.*, 2011; Townson *et al.*, 2013). The latter was written in CUDA(NVIDIA, 2011) and hence run only on Nvidia's GPU cards. In our dose calculations, we assumed that transmission through jaws was negligible and hence did not perform MC simulations inside jaws. Instead, simple geometrical tests were used to either accept the particle, if it passed through the jaws' aperture defined by their upper surfaces, or reject it otherwise. Particle transport simulation was not performed within MLC leaves in Intensity Modulated Radiation Therapy (IMRT) cases. A fluence map was generated according to the leaf sequence of a treatment plan with MLC transmission considered through a geometric model (Boyer and Li, 1997). The fluence map value at the point where the particle intersects with the fluence map was used as the weighting factor carried by this particle. We have calibrated our MC dose engine for absolute dose calculations, such that the dose at $d_{max}$ is 1 Gy under a SAD setup, when delivering 100 MU for a 10×10 cm² open field.

In our study, the dose distributions in a water phantom for five open fields, 2×2 cm², 5×5 cm², 10×10 cm², 20×20 cm² and 40×40 cm², with source-to-surface distance (SSD) set to 100 cm, were calculated with our analytical source model and the reference phase-space file, respectively. 100 MU was delivered for each case. The water phantom had a size of 60×60×60 cm³ and a resolution for $x$, $y$, and $z$ direction of 0.25×0.25×0.2 cm³. The beam normally impinges on the phantom along the z direction

We also validated our source model on one prostate IMRT patient case and one H&N IMRT patient case. There were 7 coplanar beams in the treatment plan for the prostate case. The patient model was derived from the CT image with a voxel size of 0.195×0.195×0.25 cm³. The H&N patient case had 5 coplanar beams at 0° couch angle and 1 non-coplanar beam at 90° couch angle. The voxel size used for dose calculation was 0.137×0.137×0.125cm³. The source-to-axis distances (SAD) for both cases were 100 cm. The prescription doses for these two cases were 81 Gy and 32 Gy, respectively.

## 3. Results

### *3.1 Model construction*

Fig. 5 shows the particle numbers of the primary photon PSR $NP_{pp,i}$, the scattered photon PSR $NP_{ps,i}$ and the contaminant electron PSR $NP_{e,i}$ as functions of the radius and energy of the PSRs.

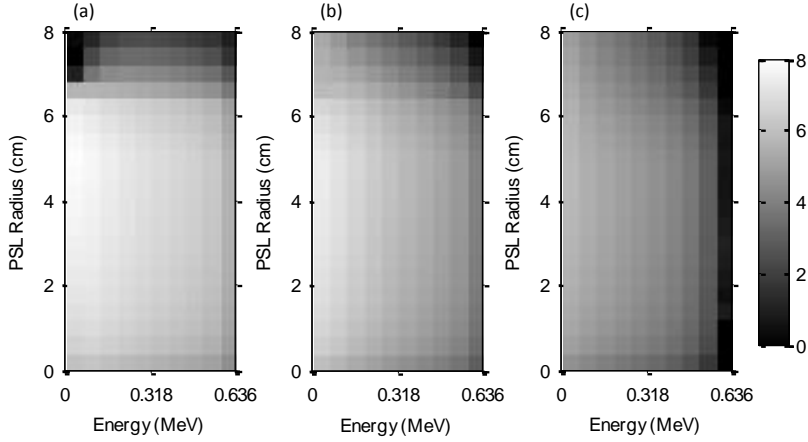

**Figure 5.** (a~c) Particle numbers of the primary photon PSR $NP_{pp,i}$, the scattered photon PSR $NP_{ps,i}$ and the contaminant electron PSR $NP_{e,i}$ are shown as functions of the radius and energy of the PSRs, respectively. Gray scale indicates common logarithm of the particle number.

The direction histograms for two representative scattered photon PSRs, $PSR_{sp}(0,0)$ (i.e. $Rbin = 0$ and $Ebin = 0$) and $PSR_{sp}(9,3)$, are shown in the two rows of Fig. 6. Each of the histogram is in fact a 2D function of $\alpha$ and $\beta$. To make presentation clear, we plot in the two columns 1D cut of the distribution along $\beta = 0$ and $\alpha = 0$, respectively. For $PSR_{sp}(0,0)$, the radius ranges in 0~0.4 cm and energy ranges in 0~0.636 MeV. The direction histogram of our source model presented by solid lines show a good match with that of the reference phase-space file denoted in dots, except at the two ends of the histogram corresponding to large scattering angles. However, the percentage of particles in this region was very small, and hence the discrepancy was expected acceptable. Another discrepancy was found at the center of the direction histogram corresponding to very small scattering angles, where a hollow was observed in the histogram. The reason for this area and our method to handle this issue has been presented in Sec 2.3.2.3. For $PSR_{sp}(9,3)$ with radius ranging in 3.6~4.0 cm and energy ranging in 1.908~2.544 MeV (around the mean energy of the beam), similar behavior was observed. Unlike the model we fitted for $PSR_{sp}(0,0)$, these three Gaussian components show different shifts in the $\alpha$ dimension, which is ascribed to the more outward scattered photon at this ring location. It could be also observed that the direction distribution was always symmetric about the $\beta$ axis due to the beam symmetry. These figures also demonstrated that three Gaussian components were enough to model the direction distribution for this scattered photon PSR.

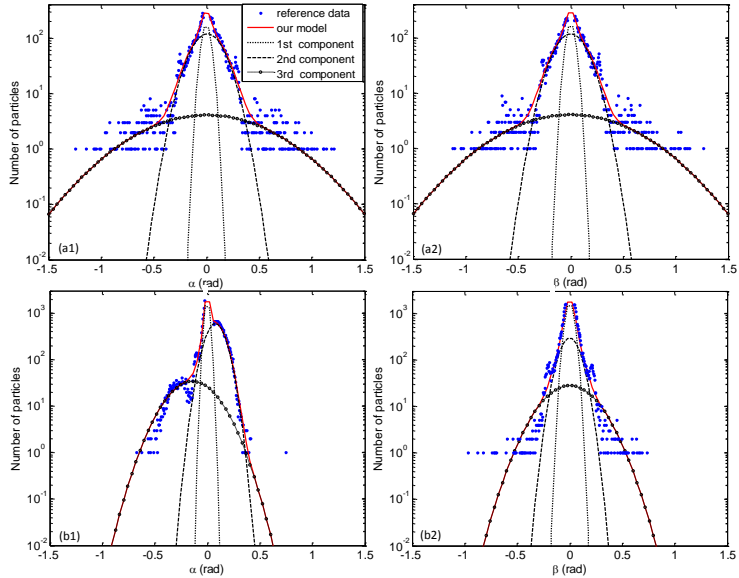

**Figure 6.** Model fitting results of particle direction distributions for scattered photon PSRs. (a1)~(a2) Particle direction histogram for $PSR_{sp}(0,0)$, showing particle number as a function of $\alpha$ and $\beta$, respectively; (b1)~(b2) Particle direction histogram for $PSR_{sp}(9,3)$. Solid circles denote the reference direction histogram obtained from the phase-space file, while the solid lines denote the fitted direction histogram using our multiple Gaussian distribution model. The dotted lines, dashed lines and dashed lines with open circles denote the first, second and third Gaussian component in our model, respectively.

## 3.1 Water phantom

With the model built, we first compared the dose distributions calculated with our analytical source model with those calculated with the reference phase-space file in five open fields with sizes 40×40 $cm^2$, 20×20 $cm^2$, 10×10 $cm^2$, 5×5 $cm^2$, 2×2 $cm^2$ in a water phantom. SSD was 100 cm in all cases. The depth dose curves and their dose differences are shown in Fig. 7. The inline and crossline lateral dose profiles at three depths 1.5 cm, 10 cm, 20 cm were compared and the results are shown in Fig. 7 and Fig. 8, respectively. Good matches were observed in both depth dose and lateral dose profiles, which demonstrated the accuracy of our analytical source model in terms of representing the reference phase-space file.

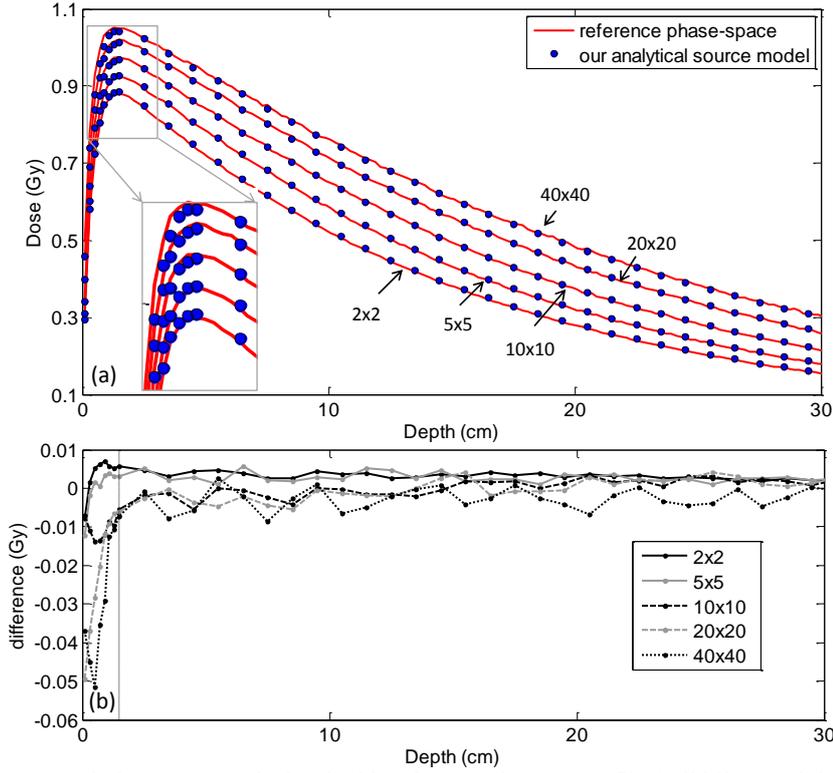

**Figure 7.** (a) Depth dose curves calculated with reference phase-space file (solid line) and those calculated with our analytical source model (dots). Five open fields (40×40 cm², 20×20 cm², 10×10 cm², 5×5 cm², 2×2 cm²) with SSD 100 cm are compared here. The build-up region was zoomed in for better display. (b) Dose differences between these two sets of the depth dose curves for the five open fields.

These dose results shown above were further quantitatively compared. Different dose comparison methods were performed based on the regions as suggested by AAPM task group 53(Fraass *et al.*, 1998). Specifically, the depth-dose curves were evaluated in two regions separately: build-up region and the region after build-up. For the build-up region which is a high-gradient region, distance-to-agreement (DTA) was employed for comparison. Given the $i_{th}$ comparison dose point in a reference dose distribution calculated with the phase-space file, $D_i^{ref}$, DTA is the distance $s$ from the nearest dose point to the corresponding dose point $D_i^{cal}$ in our dose distribution calculated with our source model, such that $D_{i+s}^{cal} = D_i^{ref}$. For the region after build-up, the root-mean-square (RMS) difference and the maximum difference were calculated as follows:

$$RMS(\%) = \frac{1}{D_{max}^{ref}} \sqrt{\frac{1}{N} \sum_{i=1}^{N} \left( D_i^{cal} - D_i^{ref} \right)^2}, \tag{13}$$

$$Max(\%) = \frac{1}{D_{max}^{ref}} max_i \left| D_i^{cal} - D_i^{ref} \right|. \tag{14}$$

Here, $D_{max}^{ref}$ denotes the depth dose value at $d_{max}$ for our reference dose distribution.

The lateral inline and crossline dose profiles were divided into three parts for evaluation: inner beam, penumbra region, and outer beam (Low *et al.*, 1998). Since the penumbra region is also a high-gradient region, DTA was adopted to evaluate the discrepancies between two sets of doses at that region. For the inner and outer beam, we still used RMS difference and maximum difference as our evaluation metrics.

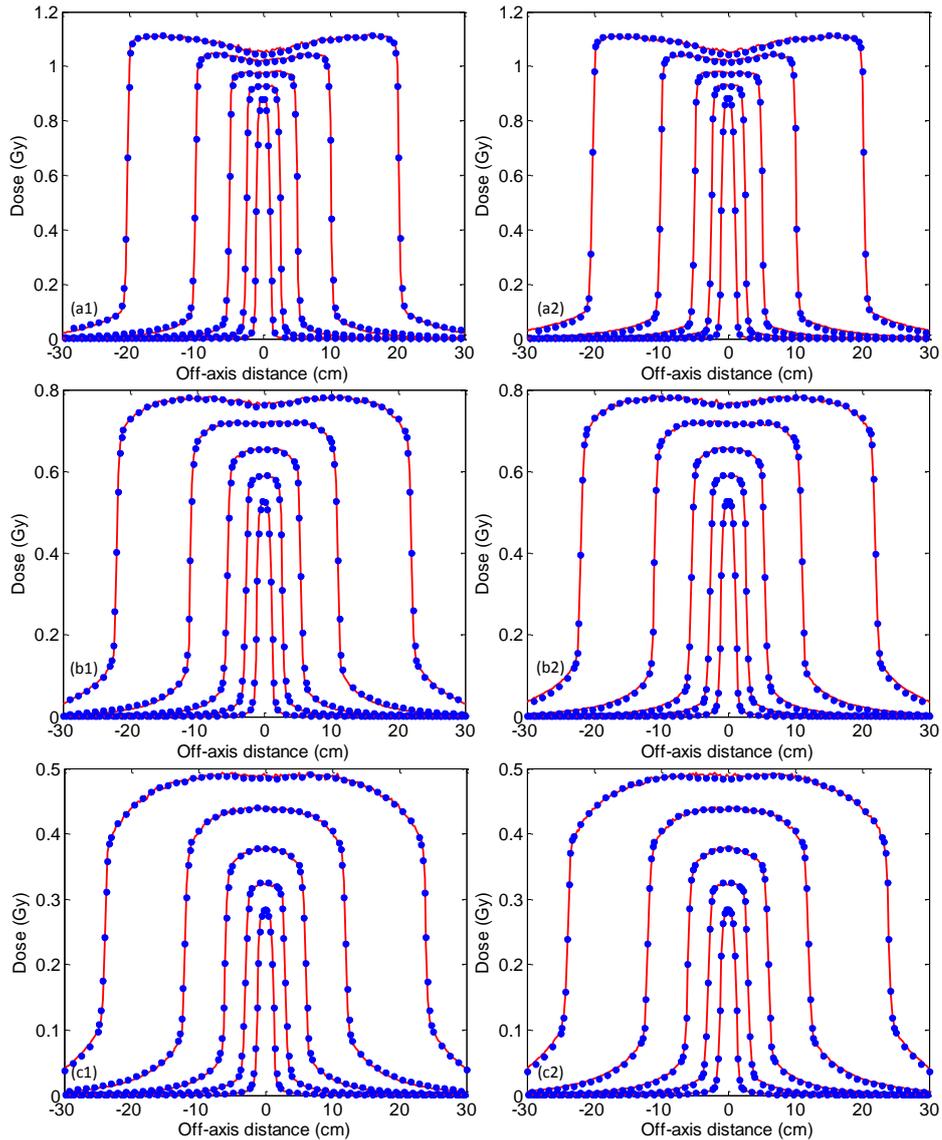

**Figure 8.** The comparison results of the inline and crossline lateral dose profiles between those calculated with the reference phase-space file (solid line) and those calculated with our analytical source model (dots). Five open fields ($40 \times 40$ $cm^2$, $20 \times 20$ $cm^2$, $10 \times 10$ $cm^2$, $5 \times 5$ $cm^2$, $2 \times 2$ $cm^2$) with SSD 100 cm are compared here. (a1~c1) show the inline dose profiles of those open fields at depth 1.5 cm, 10 cm and 20 cm, respectively. (a2~c2) show the corresponding crossline dose profiles. Same legend as in Figure 4 is used.

The comparison results of depth dose were shown in Table 1. We can see that for the build-up region of the depth dose curves for all the five open fields, the average DTA between our calculated depth dose and the reference depth dose was less than 1 mm, and the maximum DTA was 2 mm, which was clinically acceptable. After the build-up area, RMS of 0.18 ~ 0.37% and max difference of 0.52% ~ 0.93% was observed. The comparison results of inline and crossline lateral dose profiles were shown in Table 2 and Table 3, respectively. The average DTA for the penumbra regions for both inline and crossline profiles were smaller than 0.9 mm and the maximum DTA were no larger than 2 mm. For the inner beam regions, we reached a RMS smaller than 0.75% and maximal difference less than 1.8%. The RMS calculated for the outer beam regions were within 1.1% and the maximal difference within 1.2%. All these values illustrated a good match between the dose calculated with our analytical source model and the dose calculated with the reference phase-space file. This hence demonstrated the efficacy of our analytical PSR source model in terms of representing the reference phase-space file.

**Table 1.** Quantitative comparison results of depth dose curves between the dose calculated with our source model and that calculated with the reference phase-space file.

| Field size (cm$^2$) | Build-up region | | Region after build-up | |
|---|---|---|---|---|
| | Average DTA(cm) | Maximum DTA(cm) | RMS(%) | Max(%) |
| 40×40 | 0.096 | 0.202 | 0.332 | 0.934 |
| 20×20 | 0.086 | 0.185 | 0.186 | 0.529 |
| 10×10 | 0.046 | 0.126 | 0.206 | 0.733 |
| 5×5 | 0.011 | 0.036 | 0.313 | 0.596 |
| 2×2 | 0.016 | 0.033 | 0.377 | 0.682 |

**Table 2.** Quantitative comparison results of inline lateral dose profiles between the dose calculated with our source model and the dose calculated with the reference phase-space file.

| Field size (cm$^2$) | Depth (cm) | Penumbra | | Inner beam | | Outer beam | |
|---|---|---|---|---|---|---|---|
| | | Average DTA(cm) | Maximum DTA(cm) | RMS(%) | Max(%) | RMS(%) | Max(%) |
| 40×40 | 1.5 | 0.053 | 0.095 | 0.604 | 1.633 | 0.559 | 0.895 |
| | 10 | 0.056 | 0.111 | 0.358 | 0.969 | 0.210 | 0.370 |
| | 20 | 0.065 | 0.127 | 0.276 | 0.806 | 0.243 | 0.401 |
| 20×20 | 1.5 | 0.073 | 0.157 | 0.638 | 1.358 | 0.724 | 1.101 |
| | 10 | 0.059 | 0.083 | 0.279 | 0.681 | 0.448 | 0.985 |
| | 20 | 0.069 | 0.150 | 0.224 | 0.538 | 0.228 | 0.336 |
| 10×10 | 1.5 | 0.072 | 0.119 | 0.750 | 1.406 | 0.446 | 0.854 |
| | 10 | 0.069 | 0.129 | 0.273 | 0.515 | 0.257 | 0.502 |
| | 20 | 0.072 | 0.141 | 0.147 | 0.530 | 0.130 | 0.249 |
| 5×5 | 1.5 | 0.063 | 0.115 | 0.382 | 0.767 | 0.225 | 0.383 |
| | 10 | 0.061 | 0.120 | 0.226 | 0.519 | 0.117 | 0.226 |
| | 20 | 0.084 | 0.200 | 0.280 | 0.444 | 0.048 | 0.142 |
| 2×2 | 1.5 | 0.052 | 0.104 | 0.315 | 0.363 | 0.227 | 1.026 |
| | 10 | 0.063 | 0.103 | 0.434 | 0.612 | 0.069 | 0.393 |
| | 20 | 0.080 | 0.158 | 0.307 | 0.405 | 0.023 | 0.149 |

**Table 3.** Quantitative comparison results of crossline lateral dose profiles between the dose calculated with our source model and the dose calculated with the reference phase-space file.

| Field size (cm$^2$) | Depth (cm) | Penumbra | | Inner beam | | Outer beam | |
|---|---|---|---|---|---|---|---|
| | | Average DTA(cm) | Maximum DTA(cm) | RMS(%) | Max(%) | RMS(%) | Max(%) |
| 40×40 | 1.5 | 0.033 | 0.101 | 0.611 | 1.723 | 1.035 | 1.162 |
| | 10 | 0.048 | 0.185 | 0.388 | 1.035 | 0.745 | 0.877 |
| | 20 | 0.033 | 0.095 | 0.273 | 0.726 | 0.491 | 0.619 |
| 20×20 | 1.5 | 0.029 | 0.118 | 0.605 | 1.378 | 0.767 | 1.185 |
| | 10 | 0.040 | 0.147 | 0.358 | 0.848 | 0.468 | 0.707 |
| | 20 | 0.040 | 0.125 | 0.235 | 0.607 | 0.262 | 0.376 |
| 10×10 | 1.5 | 0.028 | 0.102 | 0.711 | 1.265 | 0.527 | 1.116 |
| | 10 | 0.021 | 0.069 | 0.268 | 0.695 | 0.317 | 0.678 |
| | 20 | 0.022 | 0.055 | 0.146 | 0.347 | 0.177 | 0.347 |
| 5×5 | 1.5 | 0.022 | 0.066 | 0.538 | 0.783 | 0.445 | 0.940 |
| | 10 | 0.020 | 0.056 | 0.214 | 0.499 | 0.242 | 0.525 |
| | 20 | 0.020 | 0.051 | 0.194 | 0.361 | 0.117 | 0.302 |
| 2×2 | 1.5 | 0.011 | 0.045 | 0.407 | 0.446 | 0.174 | 0.262 |
| | 10 | 0.018 | 0.076 | 0.505 | 0.612 | 0.089 | 0.160 |
| | 20 | 0.026 | 0.116 | 0.306 | 0.405 | 0.040 | 0.076 |

Finally, the output factors of our source model were also calculated and compared with those of the reference phase-space file in Fig. 9. Here, the output factors were defined to be the ratio of the central axis dose at a depth of 5 cm with SSD 95 cm for a given field to that for a $10 \times 10$ cm$^2$ open field. The maximal relative difference between these two sets of output factors is about 0.5%.

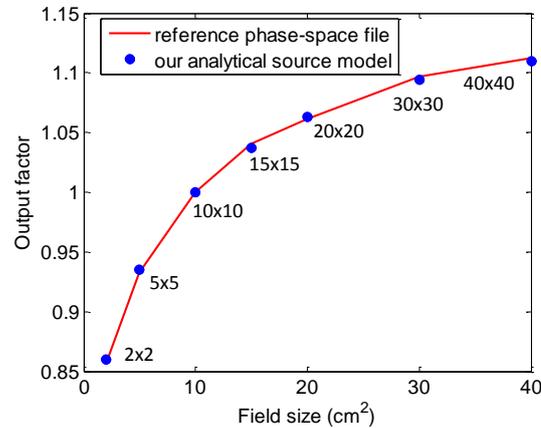

**Figure 9.** Output factors of the reference phase-space file and those of our analytical PSL-based source model.

### 3.2 Patient cases

Dose calculation results of one prostate IMRT patient case were shown in Fig. 10. The first row presented the plan dose calculated with our analytical source model. The average statistical uncertainty of the calculated dose was less than 0.5% for dose above 50% of the prescription dose. The dose distribution was shown in transverse, coronal and sagittal views, respectively, in three columns. We draw the isodose lines corresponding to 90%, 70%, 50%, 30% and 10% of prescription dose and show them in dotted lines in the second row of Fig. 10. In order to illustrate the efficacy of our source model on patient cases, the isodose lines of the dose calculated using the reference phase-space file as source model in our dose engine were also shown in Fig. 10 in solid lines for comparison. It could be observed that these two sets of isodose lines overlapped with each other at most regions. Our source model was also tested on a H&N IMRT patient case and a good match was obtained between the isodose lines of the dose distribution calculated with our source model and those of the dose calculated with the reference phase-space file as well, shown in Fig. 11. The 3D gamma index test (Low *et al.*, 1998; Gu *et al.*, 2011) with 2%/2mm criteria was performed on these two patient cases to further quantitatively evaluate the dose discrepancies between our source model and the reference phase-space file. Over 98.5% of dose points within 10% isodose lines pass the test for both cases. The experimental results of these two patient cases demonstrated again the capability of our source model to accurately represent the reference phase-space file in terms of dose calculation accuracy.

## 4. Discussion and Conclusions

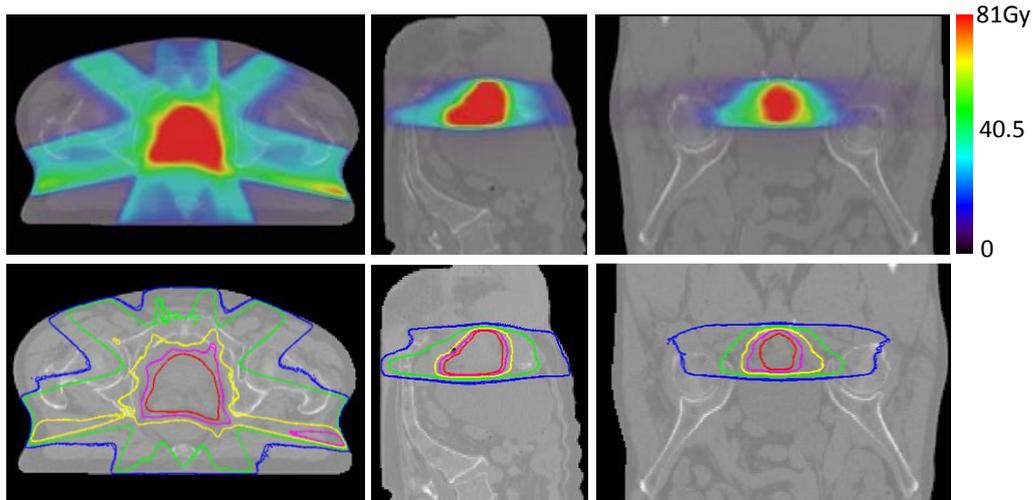

**Figure 10.** The dose distribution of the prostate IMRT patient case calculated using our PSR-based analytical source model was shown in the first row, with transverse, coronal and sagittal views in three columns, respectively. The second row shows the isodose lines of the dose distribution calculated using the reference phase-space file in PSL form (solid lines) and the isodose line of the dose calculated using our source model (dotted lines). The isodose lines corresponding to 90%, 70%, 50%, 30% and 10% of the prescription dose are drawn here.

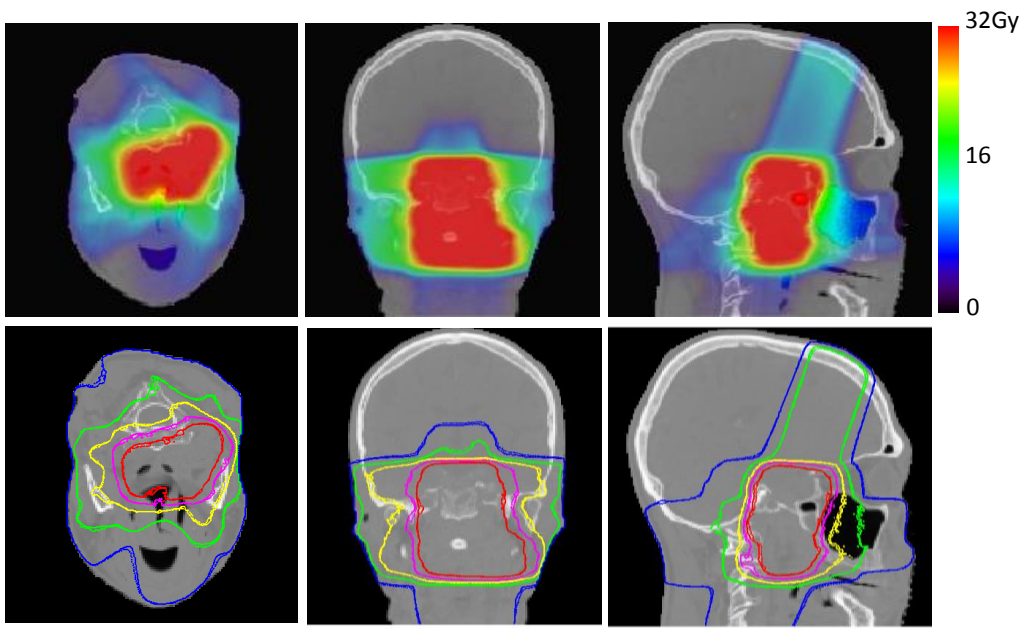

**Figure 11.** The dose distribution of the H&N IMRT patient case calculated using our PSR-based analytical source model was shown in the first row, with transverse, coronal and sagittal views in three columns, respectively. The second row shows the isodose lines of the dose distribution calculated using the reference phase-space file in PSL form (solid lines) and the isodose line of the dose calculated using our source model (dotted lines). The isodose lines corresponding to 90%, 70%, 50%, 30% and 10% of the prescription dose are drawn here.

In this paper, we have presented our PSR-based analytical source model developed for GPU-based MC dose engine. Our analytical source model consists of a set of probability density models as functions of particle location, direction and energy, along with its own relative weighting factor, built for each primary photon PSR, scattered photon PSR and electron PSR. Since the particle direction of primary photons is assumed to be pointing from the beam spot to the particle location on the phase-space plane, all the primary photon PSRs have a same direction distribution model containing a 2D Gaussian function with small standard deviations in order to model the small spot size. Multiple Gaussian components were used to model the particle direction of scattered photons, with different model parameters for each scattered photon PSR. The direction distribution of all electron PSRs was modeled as a 2D Gaussian distribution with large standard deviations to realize a moderately random and divergent distribution. We have also developed a method to analyze a phase-space file and derive corresponding model parameters. The derived model was tested against the phase-space file based source model in terms of dose calculations. The good agreements between the doses calculated using our new analytical source model and the doses using the reference phase-space file for several open fields in a water phantom and two IMRT patient cases have demonstrated the efficacy of our source model in terms of accurately representing a reference phase-space file. There are still small discrepancies between dose calculated using our analytic source model and using the reference-phase space file. This small discrepancies will be addressed later with our automatic beam commissioning model introduced in (Tian *et al.*, 2015a).

The PSR concept was proposed in this paper along with a beam symmetric assumption. This method enables us to use all the particles with in a PSR to build a source model with high statistical accuracy, which alleviated the latent variance issue of the reference phase-space file. In addition, comparing with our previous PSL-based model, the PSR approach naturally integrates the rotational symmetry in the commissioning process(Tian *et al.*, 2015a), which eliminates the needs of regularization terms in the commissioning model(Townson *et al.*, 2013; Tian *et al.*, 2014), making the computations much simpler.

For the purpose of demonstrating principle, we have only built our PSR-based analytical source model and test its efficacy and feasibility for a 6MV photon beam on a Varian linac. Our future work will extend our analytical model to photon beams of different linacs and to other energies.

**Acknowledgements**

We would also like to thank ….

**References**

Boyer A L and Li S 1997 Geometric analysis of light-field position of a multileaf collimator with curved ends *Medical Physics* **24** 757-62

Capote R 2007 IAEA nuclear and atomic data for medical applications: Phase-space database for external beam radiotherapy nuclear data for heavy charged-particle radiotherapy *Radiotherapy and Oncology* **84** S217

Davidson S, Cui J, Followill D, Ibbott G and Deasy J *Journal of Physics: Conference Series,2008),* vol. Series 102*)*: IOP Publishing) p 012004

Deng J, Jiang S B, Kapur A, Li J, Pawlicki T and Ma C 2000 Photon beam characterization and modelling for Monte Carlo treatment planning *Phys. Med. Biol.* **45** 411

Fix M K, Keall P J, Dawson K and Siebers J V 2004 Monte Carlo source model for photon beam radiotherapy: photon source characteristics *Medical physics* **31** 3106-21

Fraass B, Doppke K, Hunt M, Kutcher G, Starkschall G, Stern R and Van Dyke J 1998 American Association of Physicists in Medicine radiation therapy committee task group 53: Quality assurance for clinical radiotherapy treatment planning *Medical Physics* **25** 1773-829

Gu X, Jia X and Jiang S B 2011 GPU-based fast gamma index calculation *Physics in medicine and biology* **56** 1431

Hissoiny S, Ozell B, Bouchard H and Després P 2011 GPUMCD: A new GPU-oriented Monte Carlo dose calculation platform *Medical physics* **38** 754-64

Jahnke L, Fleckenstein J, Wenz F and Hesser J 2012 GMC: a GPU implementation of a Monte Carlo dose calculation based on Geant4 *Phys. Med. Biol.* **57** 1217

Jia X, Gu X, Graves Y J, Folkerts M and Jiang S B 2011 GPU-based fast Monte Carlo simulation for radiotherapy dose calculation *Phys Med Biol* **56** 7017-31

Jia X, Gu X, Sempau J, Choi D, Majumdar A and Jiang S B 2010 Development of a GPU-based Monte Carlo dose calculation code for coupled electron-photon transport *Phys Med Biol* **55** 3077

Jia X, Ziegenhein P and Jiang S B 2014 GPU-based high-performance computing for radiation therapy *Phys Med Biol* **59** R151-82

Jiang S B, Deng J, Li J, Pawlicki T, Boyer A L and Ma C M *The 13th international conference on the use of computers in radiotherapy, (Heidelberg, Germany, 2000),* vol. Series*)* p 434

Keall P J, Siebers J V, Jeraj R and Mohan R 2000 The effect of dose calculation uncertainty on the evaluation of radiotherapy plans *Medical Physics* **27** 478-84

Khronos OpenCL Working Group 2013 The open standard for parallel programming of heterogeneous systems. ed A Munshi

Low D A, Harms W B, Mutic S and Purdy J A 1998 A technique for the quantitative evaluation of dose distributions *Medical physics* **25** 656-61

Ma C-M 1998 Characterization of computer simulated radiotherapy beams for Monte-Carlo treatment planning *Radiation Physics and Chemistry* **53** 329-44

Ma C, Faddegon B, Rogers D and Mackie T 1997 Accurate characterization of Monte Carlo calculated electron beams for radiotherapy *Medical physics* **24** 401-16

Moré J J 1978 *Numerical analysis*: Springer) pp 105-16

NVIDIA 2011 *NVIDIA CUDA Compute Unified Device Architecture, Programming Guide, 4.0*

Rogers D W O and Mohan R *The Use of Computers in Radiation Therapy: 13th International Conference, (Heidelberg, Germany, 2000),* vol. Series*)* ed W Schlegel and T Bortfeld pp 120-2

Tian Z, Folkerts M, Li Y, Shi F, Jiang B S and Jia X 2015a An analytic linear accelerator source model for Monte Carlo dose calculations. II. Its utilization in a GPU-based Monte Carlo package and automatic source commissioning *Physics in Medicine and Biology* **(submitted)**

Tian Z, Graves Y J, Jia X and Jiang S 2014 Automatic Commissioning of a GPU-based Monte Carlo Radiation Dose Calculation Code for Photon Radiotherapy *Phys Med Biol* **59** 6467-86

Tian Z, Shi F, Folkerts M, Qin N, Jiang B S and Jia X 2015b An OpenCL-based Monte Carlo dose calculation engine (oclMC) for coupled photon-electron transport *Physics in Medicine and Biology* **(submitted)**

Townson R W, Jia X, Tian Z, Graves Y J, Zavgorodni S and Jiang S B 2013 GPU-based Monte Carlo radiotherapy dose calculation using phase-space sources *Phys Med Biol* **58** 4341-56

Verhaegen F and Seuntjens J 2003 Monte Carlo modelling of external radiotherapy photon beams *Physics in medicine and biology* **48** R107